\documentclass[12pt]{article}

\usepackage{graphicx,epstopdf,amssymb,amsfonts,amsmath,amsthm,array,
mathrsfs,amscd}

\newcommand{\pbs}[1]{\let\temp=\\#1\let\\=\temp}
\numberwithin{equation}{section}
\def\be{\begin{equation}}\def\ee{\end{equation}}
\def\cvp{\raise 2pt\hbox{,}}

  \def\im{\mathop{\rm Im}\nolimits}
\def\re{\mathop{\rm Re}\nolimits}  

 \def\d{{\rm d}}

\def\Sgrav{S_{\text{grav}}}\def\SL{S_{\text L}}\def\SM{S_{\text M}}
\def\SAY{S_{\text{AY}}}

\def\sigh{\Sigma_{h}}
\def\Zmat{Z_{\text{mat}}}\def\Zgh{Z_{\text{gh}}}
\def\Zmgh{Z_{\text{mat}+\text{gh}}}\def\Zgrav{Z_{\text{grav}}}
\def\Smat{S_{\text{mat}}}

\def\ish{\int_{\Sigma_{h}}\!\d^{2}x\sqrt{g}\,}
\def\ishbis{\int_{\Sigma_{h}}\!\d^{2}x\sqrt{g_{0}}\,}
\def\k{\boldsymbol{k}}\def\x{\boldsymbol{x}}

\theoremstyle{plain}

\theoremstyle{definition}
\theoremstyle{remark}

\def\cmp#1#2#3{{\it Comm.\ Math.\ Phys.\ }{\bf #1} (#2) #3}

\def\imath#1#2#3{{\it Invent math }{\bf #1} (#2) #3}
\def\jdiffgeo#1#2#3{{\it J.\ Diff.\ Geom.\ }{\bf #1} (#2) #3}
\begin{document}
%
%
{\pagestyle{empty}
\parskip 0in

\

\vfill
\begin{center}
{\LARGE Gravitational Actions in Two Dimensions}

\bigskip

{\LARGE and the Mabuchi Functional}

\vspace{0.4in}

Frank F{\scshape errari},$^{1}$ Semyon K{\scshape levtsov}$^{1}${\renewcommand{\thefootnote}{$\!\!\dagger$}
\footnote{On leave of absence from ITEP, Moscow, Russia.}} and 
Steve Z{\scshape elditch}$^{2}$
\\
\medskip
{\it $^{1}$Service de Physique Th\'eorique et Math\'ematique\\
Universit\'e Libre de Bruxelles and International Solvay Institutes\\
Campus de la Plaine, CP 231, B-1050 Bruxelles, Belgique}

\smallskip

{\it $^{2}$Department of Mathematics, Northwestern University\\
Evanston, IL 60208, USA}

\smallskip

{\tt frank.ferrari@ulb.ac.be, semyon.klevtsov@ulb.ac.be, 
zelditch@math.northwestern.edu }

\medskip

\end{center}
\vfill\noindent

The Mabuchi energy is an interesting geometric functional on the space of K\"ahler metrics that plays a crucial r\^ole in the study of the geometry of K\"ahler manifolds. We show that this functional, as well as other related geometric actions, contribute to the effective gravitational action when a massive scalar field is coupled to gravity in two dimensions in a small mass expansion. This yields new theories of two-dimensional quantum gravity generalizing the standard Liouville models.

\vfill

\medskip
%
\begin{flushleft}
\today
\end{flushleft}
\newpage\pagestyle{plain}
\baselineskip 16pt
\setcounter{footnote}{0}

}

\section{Introduction}
\label{IntroSec}

When a matter quantum field theory is coupled to gravity, the metric dependence of the partition function induces an effective gravitational action. On a two-dimensional euclidean space-time, when the quantum field theory is conformally invariant, this effective action is always proportional to the Liouville action \cite{polyakov}. In the so-called conformal gauge, for which the metric $g$ is proportional to a fixed background metric $g_{0}$,
\begin{equation}
\label{confgauge}
g=e^{2\sigma}g_{0}\, ,
\end{equation}
the Liouville action takes the simple form
\be\label{Liouvilleaction} \SL(g_{0},g) = \int\!\d^{2}x\sqrt{g_{0}}\,\bigl( g_{0}^{ab}\partial_{a}\sigma\partial_{b}\sigma + R_{0}\sigma\bigr)\, ,\ee
where $R_{0}$ is the Ricci scalar for the metric $g_{0}$. The resulting Liouville models for two-dimensional quantum gravity have been extensively studied in the literature.

The gravitational coupling of non-conformally invariant matter has been much less studied (see e.g.\ \cite{nonconf}). In this case, one expects a complicated and model-dependent gravitational action. It is unclear a priori if new interesting and simple actions, yielding models with nice geometrical and physical features, can emerge.

Recently, the authors have explained 
that two-dimensional quantum gravity could be naturally formulated in the context of K\"ahler geometry \cite{FKZ1,FKZ2}. In this context,
mathematicians have much used and studied natural functionals
defined on the space of K\"ahler potentials, the most salient being 
the Mabuchi action \cite{Mabuchi}, also called the Mabuchi energy in the mathematical literature. Our aim in the present 
work is to investigate whether such actions could contribute to the 
gravitational effective action in some simple physical models. We shall 
see that they do, yielding new models of two-dimensional quantum gravity with potentially very rich mathematical and physical features.

Our space-time will always be a compact Riemann surface $\Sigma_{h}$ of genus $h$. An arbitrary riemannian metric on $\Sigma_{h}$ can be
parametrized by a finite set of complex moduli $\tau$ and the conformal factor $e^{2\sigma}$, with a background metric $g_{0}(\tau)$ in \eqref{confgauge} of the form $g_{0}= 2g_{0z\bar z}|\d z|^{2}$.
In the K\"ahler framework, we write the conformal factor $e^{2\sigma}$ in terms of the area $A$ (which is the unique 
K\"ahler modulus in two dimensions) and the K\"ahler potential $\phi$,
\be\label{phidef} e^{2\sigma} = \frac{A}{A_{0}} - \frac{1}{2}A\Delta_{0}\phi\, ,\ee
where $A_{0}$ and $\Delta_{0}$ are the area and the positive laplacian for the metric $g_{0}$ respectively. The $\partial\bar\partial$ lemma shows that the differential equation \eqref{phidef} can always be solved for $A$ and $\phi$ in terms of $\sigma$, and the solution is unique up to constant shifts in $\phi$. 

There are many reasons to use the K\"ahler potential $\phi$ in the context of two-dimensional quantum gravity. 
In particular it is possible to rigorously define in terms of 
$\phi$ a regularized version of the gravitational path integrals \cite{FKZ1,FKZ2}. For our present purposes, 
the variable $\phi$ is useful because interesting functionals of the metric are most naturally written down in terms of $\phi$.
A famous example is the Mabuchi action \cite{Mabuchi},
\be\label{Mabdef1}
\SM (g_{0},g)  =
\ishbis\biggl[ -2\pi (1-h)\phi\Delta_{0}\phi + \Bigl(\frac{8\pi(1-h)}{A_{0}}-R_{0}\Bigr)\phi + \frac{4}{A}\sigma e^{2\sigma}\biggr]\, .\ee
This action and its higher dimensional generalizations play a central
r\^ole in K\"ahler geometry. It is well defined on the space or metrics, because it is unchanged if $\phi$ is shifted by a constant.
It is bounded from below and convex, which makes it a good candidate for an action to be used in a path integral. It satisfies the so-called cocycle conditions,
\begin{align}\label{cocM1} \SM(g_{1},g_{2}) & = -\SM(g_{2},g_{1})\, ,\\
\label{cocM2} \SM(g_{1},g_{3}) &= \SM(g_{1},g_{2}) + \SM
(g_{2},g_{3})\, ,\end{align}
which can be checked straightforwardly from the definition \eqref{Mabdef1}. The same cocycle identities are also satisfied by the Liouville action. As we shall review below, these identities actually are fundamental consistency conditions that any effective gravitational action must satisfy. Finally, the critical points of the Mabuchi action are the metrics of constant scalar curvature, another property shared in two dimensions with the Liouville action. This property is also valid for the higher dimensional generalizations of \eqref{Mabdef1}, which goes a long way in explaining 
the central r\^ole played by the Mabuchi action in the study of such metrics on general K\"ahler manifolds.
For more details and references on the profound geometrical properties of the space of metrics on a K\"ahler manifold, we refer the reader to 
\cite{FKZ2, Mabuchi2, Chen} and the comprehensive recent review \cite{PS}.

We focus in the following on the simple model of a massive scalar field $X$  with action 
\be\label{dilatongen} \Smat(X,g;q,m) = \frac{1}{8\pi} \int\!\d^{2}x\, 
\sqrt{g}\,\bigl( g^{ab}\partial_{a}X\partial_{b}X + q R X+ m^{2}
X^{2}\bigr)\, .\ee
Our main result is to show that the Mabuchi action and other simple 
functionals of the K\"ahler potential $\phi$, like the so-called Aubin-Yau action, contribute to the gravitational effective action for this model in a small mass expansion. We also study the gravitational dressing of operators of the form
\be\label{opalpha} \mathscr O_{k} = e^{k X}\ee
and show that it involves the Aubin-Yau action on top of the familiar dressing factors found in the conformal field theory limit. 

A striking feature is that it is always possible to cancel out the familiar Liouville term in the effective action, for 
example by coupling with a suitable spectator conformal field theory.
The resulting pure Mabuchi theories are entirely new and intriguing two-dimensional quantum gravity models.

\noindent\emph{Plan of the paper}

We start in Section 2 with a brief review on two-dimensional quantum gravity, gravitational effective actions and gravitational dressing.
In Section 3, we discuss the basic properties of various actions, in particular Mabuchi's and Aubin-Yau's. We derive fundamental identities relating these actions to the variations of various functionals associated with the Laplace operator. In Section 4, we apply the results of Section 3 to compute the gravitational effective action and the gravitational dressing of the operators $e^{kX}$ for the model \eqref{dilatongen}. We also compute the trace of the stress-energy tensor. Finally, in Section 5, we conclude and discuss possible extensions of our work. We have tried to make the presentation as elementary and self-contained as possible.
In particular we have included two Appendices containing simple derivations of results used in the main text.

\noindent\emph{Notations}

To a metric $g = 2g_{z\bar z}|\d z|^{2}$ on $\Sigma_{h}$ we associate its K\"ahler form,
\be\label{Kfdef} \omega = i g_{z\bar z}\d z\wedge\d\bar z\, .\ee
In two dimensions, the K\"ahler form coincides with the volume form and thus in particular the area is given by
\be\label{area} A(g) = \int_{\Sigma_{h}}\omega\, .\ee
The positive laplacian for the metric $g$, whose determinant we also denote by $g$, is defined as usual by
\be\label{deflap} \Delta f = -\frac{1}{\sqrt{g}}\partial_{a}\bigl(
\sqrt{g}g^{ab}\partial_{b}f\bigr)\, .\ee
In terms of the standard Dolbeault operators $\partial$ and $\bar\partial$ we have
\be\label{idenlap} \partial\bar\partial f = \frac{i}{2}\Delta f\,\omega\, ,\ee
from which it can be seen that the relation \eqref{phidef} is equivalent to
\be\label{phidef2} \omega = \frac{A}{A_{0}}\omega_{0}
 + i A \partial\bar\partial\phi\, .\ee
The space of K\"ahler potentials $\phi$ is simply
\be\label{Kdef}\mathscr K_{\omega_{0}} = \bigl\{ \phi :\Sigma\rightarrow \mathbb R 
\mid A_{0}\Delta_{0}\phi < 2\bigr\}\, ,
\ee
the condition on $\phi$ ensuring the strict positivity of the metrics, 
$g_{z\bar z}>0$. The scalar curvature, or Ricci scalar $R$, is such that
\be\label{Riccidef} R\,\omega = -i\partial\bar\partial\ln g\, .\ee
The integral of the scalar curvature is a topological invariant,
\be\label{euler} \int_{\Sigma_{h}} R\,\omega = 8\pi(1-h)\, .\ee
This is a useful formula, from which for instance the invariance of \eqref{Mabdef1} under constant shifts of $\phi$ is derived. Finally, the laplacians $\Delta$ and $\Delta_{0}$ and scalar curvatures $R$ and $R_{0}$ of two metrics $g$ and $g_{0}$ related by a formula of the form \eqref{confgauge} are themselves linked by the simple relations
\begin{align}\label{laprel} e^{2\sigma}\Delta
& = \Delta_{0}\, ,\\
\label{Rrelconf} e^{2\sigma} R & = R_{0} + 2 \Delta_{0}\sigma\, .
\end{align}
\section{On gravitational effective actions and dressing}
\label{Sgravsec}
\subsection{Basic definitions}
\label{Sgravdefsec}

\subsubsection{Gravitational effective actions}

The partition function $\Zmat$ of a matter quantum field theory, defined on the two-dimensional euclidean space-time $\Sigma_{h}$ endowed with a fixed metric $g$, with fields $X$, couplings $\lambda$ and action 
$\Smat (X;\lambda;g)$, is defined by the QFT path integral
\be\label{matZdef} \Zmat (\lambda;g) = \int\!\mathscr D X\,  
e^{-\Smat (X;\lambda;g)}\, .\ee
The quantum gravity partition function $Z$ is obtained by integrating further over the space of metrics on $\Sigma_{h}$,
\be\label{pi1} Z(\mu,\lambda)=\int\!\mathscr D g\, \Zmat(\lambda;g) 
e^{-\mu A(g)}\, ,
\ee
where $\mu$ is the bare cosmological constant. It is also interesting to consider the partition function $Z_{A}$ at fixed area, such that
\begin{align}\label{pi2} Z_{A}(\lambda) &= \int\!\mathscr D g\, \Zmat(\lambda;g) \delta\Bigl(\int_{\sigh}\!\d^{2}x\sqrt{g} - A\Bigr)\, ,\\
\label{Zform} Z(\mu,\lambda) &= \int\!\d A\, e^{-\mu A}Z_{A}(\lambda)\, .
\end{align}
The above formal path integrals over metrics are of course infinite because of diffeormorphism invariance and the fact that the diffeomorphism group has infinite volume. As is well-known, the effect of fixing the gauge by imposing \eqref{confgauge} is to replace 
\be\label{DgtoDgh} \mathscr D g \rightarrow \mathscr D\sigma\, \Zgh(g)\, ,\ee
where the volume form $\mathscr D\sigma$ is associated with the curved riemannian metric
\be
\label{Calabimetric}
\lVert\delta\sigma\rVert^{2} = \int_{\Sigma}\!\d^{2}x\sqrt{g}\,(\delta\sigma)^{2}=\int_{\Sigma}\!\d^{2}x\sqrt{g_{0}}\,e^{2\sigma}(\delta\sigma)^{2}\ee
on the space of conformal factors and $\Zgh(g)$ is the partition function of a ghost conformal field theory of central charge $c_{\text{gh}}=-26$ corresponding to the Faddeev-Popov determinant. 
We end up with
\be\label{Zgdef} Z(\mu,\lambda) = \int\!\mathscr D\sigma\, e^{-\mu A(g)}\Zmgh(\lambda;g)\, ,
\ee
where
\be\label{Zmghdef} 
\Zmgh(\lambda;g)=\Zmat(\lambda;g)\times\Zgh(g)
\ee
the partition function of the matter plus ghost QFT.
A regularization of the path integral over $\sigma$ and in particular a rigorous construction of the volume form $\mathscr D\sigma$ is given in 
\cite{FKZ1,FKZ2}.

The \emph{gravitational effective action,} or simply the gravitational action, is defined by
\be\label{Sgravdef}\Sgrav(g_{0},g;\lambda) = 
-\ln\frac{\Zmgh(\lambda,g)}{\Zmgh(\lambda,g_{0})}\,\cdotp\ee
It depends on the metric $g$, on a reference (or background) metric $g_{0}$ and on the couplings $\lambda$ in the matter QFT. In terms of 
$\Sgrav$ the partition function reads
\begin{align}\label{ZSgrav} Z(\mu,\lambda)& = \Zmgh(\lambda;g_{0})\int\!\mathscr D\sigma\, e^{-\mu A(g)-\Sgrav(g_{0},g;\lambda)}\\
\label{ZSgravprod}
& = \Zmgh(\lambda;g_{0})\times\Zgrav (\mu,\lambda;g_{0})
\end{align}
where the gravitational partition function is defined by
\be\label{Zgravdef} \Zgrav(\mu,\lambda;g_{0}) = 
\int\!\mathscr D\sigma\, e^{-\mu A(g)-\Sgrav(g_{0},g;\lambda)}\, .
\ee
The calculation of the quantum gravity partition function can thus be split into two parts. The first part is a standard QFT calculation in a \emph{fixed} background metric yielding $\Zmgh$. The second part is
a gravitational partition function governed by the gravitational action
$\Sgrav$. 

\subsubsection{Gravitational dressing}

The above discussion can be repeated straightforwardly when operators are inserted in the path integral. For example, correlators of scalar operators in the matter QFT are defined by
\be\label{matcorrdef} \bigl\langle \mathscr O_{1}(x_{1})\cdots
\mathscr O_{n}(x_{n})\bigr\rangle_{\text{mat},\, g} = \frac{1}{\Zmat (\lambda;g)}
\int\!\mathscr D X\,  \mathscr O_{1}(x_{1})\cdots
\mathscr O_{n}(x_{n})\, e^{-\Smat (X;\lambda;g)}\, .\ee
The integrated version of these correlators
\be\label{mticdef}\int_{\sigh}\!\d^{2}x_{1}\sqrt{g(x_{1})}\cdots 
\int_{\sigh}\!\d^{2}x_{n}\sqrt{g(x_{n})}\,
\bigl\langle \mathscr O_{1}(x_{1})\cdots
\mathscr O_{n}(x_{n})\bigr\rangle_{\text{mat},\, g}
\ee
is diffeomorphism invariant and can be averaged over metrics, yielding the quantum gravity correlators. If we introduce the \emph{gravitational
dressing} of the QFT correlators, defined by
\be\label{dressingdef}
D_{\mathscr O_{1}\cdots\mathscr O_{n}}(x_{1},\ldots,x_{n};
\lambda; g_{0},g) = e^{2\sum_{i=1}^{n}\sigma(x_{i})}
\frac{\bigl\langle \mathscr O_{1}(x_{1})\cdots
\mathscr O_{n}(x_{n})\bigr\rangle_{\text{mat},\, g}}{\bigl\langle \mathscr O_{1}(x_{1})\cdots
\mathscr O_{n}(x_{n})\bigr\rangle_{\text{mat},\, g_{0}}}\,\cvp\ee
then we can cast the quantum gravity correlators in the form
\begin{multline}\label{corrform1} \bigl\langle\mathscr O_{1}\cdots
\mathscr O_{n}\bigr\rangle=
\int_{\sigh}\!\d^{2}x_{1}\sqrt{g_{0}(x_{1})}\cdots 
\int_{\sigh}\!\d^{2}x_{n}\sqrt{g_{0}(x_{n})}\\
\bigl\langle \mathscr O_{1}(x_{1})\cdots
\mathscr O_{n}(x_{n}) \bigr\rangle_{\text{mat},\, g_{0}}
\bigl\langle D_{\mathscr O_{1}\cdots\mathscr O_{n}}(x_{1},\ldots,x_{n};
\lambda; g_{0},g)\bigr\rangle_{\text{grav}}
\, .
\end{multline}
As for the case of the partition function, the calculation of the correlator splits into two parts, a standard QFT calculation yielding the
matter QFT correlator and the calculation of a correlator
\be\label{gravexpdef} \bigl\langle D_{\mathscr O_{1}\cdots\mathscr O_{n}}
\bigr\rangle = \frac{1}{\Zgrav(\mu,\lambda;g_{0})}
\int\!\mathscr D\sigma\, D_{\mathscr O_{1}\cdots\mathscr O_{n}}
e^{-\mu A(g)-\Sgrav(g_{0},g;\lambda)}
\ee
in the effective gravitational theory governed by $\Sgrav$.

\subsubsection{The case of a conformal field theory}

In the case of a matter CFT of central charge $c$, the gravitational action is simply \cite{polyakov}
\be\label{sgravCFT} \Sgrav(g_{0},g;c)=\frac{26-c}{24\pi}\SL(g_{0},g)\, , 
\ee
where $\SL$ is the Liouville action \eqref{Liouvilleaction}. Note that $c-26$ is the total central charge of the matter plus ghost system.
The gravitational dressing can also be exactly determined for CFT operators. For example, if the operators $\mathscr O_{i}$, $1\leq i\leq n$, are conformal primaries of conformal dimensions $\Delta_{i}$ (the case of a marginal operator corresponding to $\Delta = 2$), then the dressing is simply
\be\label{dressingCFT}
D_{\mathscr O_{1}\cdots\mathscr O_{n}}(x_{1},\ldots,x_{n};
\lambda; g_{0},g) = e^{\sum_{i=1}^{n}(2-\Delta_{i})\sigma(x_{i})}\, .
\ee
\subsection{Consistency conditions}
\label{cocysec}

Any model (associated with a general matter QFT, not necessarily conformal) must satisfy two basic consistency requirements: background independence and the cocycle identities.

\subsubsection{Background independence}

Background independence is simply the statement that the quantum gravity theory does not depend on the choice of background metric $g_{0}$ in
\eqref{confgauge}. For example, the partition function $Z$ in \eqref{ZSgrav} does not depend on $g_{0}$ and thus the $g_{0}$ dependence must cancel on the right-hand side of \eqref{ZSgravprod} between $\Zmgh$ and $\Zgrav$. Equivalently, the total matter plus ghost plus gravitational system must be a \emph{conformal field theory of central charge zero}. 
Note that this is true even
if the matter QFT is not conformal. Similarly, since the left-hand side of \eqref{corrform1} is independent of the reference metric $g_{0}$, the same must be true for the right-hand side.

\subsubsection{Cocycle identities}

Any gravitational effective action must satisfy cocycle identities that follow immediately from the definition \eqref{Sgravdef},
\begin{align}\label{cocycle1} \Sgrav(g_{1},g_{2}) & = -\Sgrav(g_{2},g_{1})\, ,\\
\label{cocycle2} \Sgrav(g_{1},g_{3}) &= \Sgrav(g_{1},g_{2}) + \Sgrav
(g_{2},g_{3})\, .\end{align}
It is instructive to check by hand that the Liouville action \eqref{Liouvilleaction} and the Mabuchi action \eqref{Mabdef1} both satisfy these identities. We shall build other functionals in the next Section that also satisfy the same constraints.

The identity \eqref{cocycle2} has a useful differential formulation. In general, let us consider any functional $S(g_{1},g_{2})$ satisfying
\be\label{cocgeneral} S(g_{1},g_{3})=S(g_{1},g_{2})+S(g_{2},g_{3})\, .\ee
Taking the functional derivative of \eqref{cocgeneral} with respect to $g_{3}$ we get
\be\label{cocdiff1} \frac{\delta S(g_{1},g_{3})}{\delta g_{3}(x)} =
\frac{\delta S (g_{2},g_{3})}{\delta g_{3}(x)}\,\cdotp\ee
This being valid for any $g_{1}$ and $g_{2}$, we conclude that
\be\label{funcdiffcoc} 
\frac{\delta S(g_{0},g)}{\delta g_{ab}(x)} = \frac{1}{4\pi} t^{ab}(x;g)\ee
depends only on $g$ and not on $g_{0}$ (the factor $1/(4\pi)$ in \eqref{funcdiffcoc} is conventional). 
Since we shall always be in the gauge \eqref{confgauge}, the variation of the metric is given by
\be\label{confvarmet}\delta g_{ab} = 2g_{ab}\delta\sigma\ee
and the variation of $S$ can thus be written 
\be\label{deltaS2} \delta_{g}S (g_{0},g) = \frac{1}{2\pi}\int_{\sigh}\!\d^{2}x\sqrt{g(x)}\, t(x;g)\delta \sigma(x)\, ,\ee
in terms of the trace $t = g_{ab}t^{ab}=t^{a}_{\ a}$ that depends on $g$ but not on $g_{0}$. Equivalently, from \eqref{phidef} we see that a variation of $\sigma$ is equivalent to a variation of $A$ and $\phi$ related by
\be\label{dsigdphi}\delta\sigma = \frac{\delta A}{2A} - \frac{1}{4} A\Delta\delta\phi\, ,\ee
and thus equation \eqref{deltaS2} is equivalent to
\be\label{dSphi} \delta_{g}S(g_{0},g) = \frac{\delta A}{4\pi A}\ish t
-\frac{A}{8\pi}\ish\Delta t\delta\phi\, .
\ee

Conversely, if any of the equivalent conditions \eqref{funcdiffcoc}, 
\eqref{deltaS2} or \eqref{dSphi} is satisfied, then
\be\label{solvecocycle} S(g_{0},g) = \Psi(g) - \Psi(g_{0})\ee
for a primitive $\Psi$ of $t$ with respect to $g$. Such a functional $S$ automatically satisfies \eqref{cocycle2} and $S(g_{0},g)=-S(g,g_{0})$ as well.

When $S$ is not an arbitrary abstract functional satisfying \eqref{cocgeneral} but corresponds to a gravitational effective action, the definition \eqref{Sgravdef} shows 
that 
\be\label{PsiZ}\Psi(g) = -\ln\Zmgh(g)\ee
and thus
\be\label{tTrel} t^{ab}(g) = \bigl\langle T^{ab}
\bigr\rangle_{\text{mat}+\text{gh}}
\ee
is the vacuum expectation value of the stress-energy tensor $T^{ab}$ of the matter plus ghost QFT in the fixed background geometry $g$. The gravitational effective action is thus always determined by the expectation value of the trace of the stress-energy tensor.

For example, if the matter theory is a CFT, the trace of the stress-energy tensor is zero classically and its possible non-zero value at the quantum level is entirely due to a quantum anomaly. An anomaly must always be a local functional and the only possibility for the trace anomaly in two dimensions consistent with dimensional analysis and diffeomorphism invariance is $t = a + b R$ for some constants $a$ and $b$. The constant $a$ can always be absorbed in a redefinition of the cosmological constant and thus can be discarded. From \eqref{deltaLiouville}, the gravitational effective action must be the Liouville action, as indicated in \eqref{sgravCFT}. The precise coefficient $b=(26-c)/(24\pi)$ can be computed in various ways, for example in the flat space limit. We shall generalize this reasoning for the non-conformal model \eqref{dilatongen} in Section \ref{stresssec}.

Finally, let us note that the gravitational dressing factors satisfy exponentiated versions of the cocycle conditions that follow immediately from their definition
\eqref{dressingdef},
\begin{align}\label{cocycle3}
D_{\mathscr O_{1}\cdots\mathscr O_{n}} (g_{1},g_{2}) & =
\frac{1}{D_{\mathscr O_{1}\cdots\mathscr O_{n}}(g_{2},g_{1})}\, \cvp\\
\label{cocycle4}
D_{\mathscr O_{1}\cdots\mathscr O_{n}}(g_{1},g_{3}) &= D_{\mathscr O_{1}\cdots\mathscr O_{n}}(g_{1},g_{2})D_{\mathscr O_{1}\cdots\mathscr O_{n}}(g_{2},g_{3})\, .
\end{align}
\section{Functionals and variations}
\label{MAYsec}

We are now going to derive a set identities for the
variation under change of metric of various functionals. All these  identities will then be applied in Section 4 to compute the gravitational effective action and gravitational dressing of the model \eqref{dilatongen}. The technical derivations that we present below and in the Appendices are not necessary in understanding Section 4 and thus may be skipped on a first reading. 

\subsection{Building blocks for gravitational actions}
\label{bbsec}
\subsubsection{The area functional}

The simplest functional that can enter in a gravitational effective
action is the area, or more generally any function of the area,
\be\label{Area} S_{f}(g_{0},g) = f(A) - f(A_{0})\, .\ee
This is associated with a term
\be\label{tracearea} t_{f} = 4\pi f'(A)\ee
in the trace of the stress-energy tensor.

\subsubsection{The Liouville action}

The variation of the Liouville action \eqref{Liouvilleaction} with respect to $\sigma$,
\be\label{deltaLiouville} 
\delta \SL(g_{0},g) = \int_{\sigh}\!\d^{2}x\sqrt{g}\, R\delta\sigma\, .
\ee
yields the term
\be\label{traceLiou} t_{\text L}= 2\pi R\ee
in the trace of the stress-energy tensor.

\subsubsection{The Mabuchi action}

The Mabuchi action, defined in \eqref{Mabdef1}, can also be written as
\begin{align}\label{SMdef2} \SM (g_{0},g) 
& = \int_{\sigh}\biggl[ -4i\pi (1-h) \partial\phi\wedge\bar\partial
\phi + \Bigl(\frac{8\pi(1-h)}{A_{0}}-R_{0}\Bigr)\phi\omega_{0} +
\frac{2}{A}\omega\ln\frac{\omega}{\omega_{0}}\biggr]
\\\label{SMdef3}
&=\int_{\sigh}\biggl[ 4\pi(1-h) \phi\Bigl(
\frac{\omega_{0}}{A_{0}} + \frac{\omega}{A}\Bigr) - \phi R_{0}\omega_{0}
+ \frac{2}{A}\omega\ln\frac{\omega}{\omega_{0}}\biggr]\, .
\end{align}
Let us note that this definition involves metrics $g_{0}$ and $g$ that do not necessarily have the same area. In the mathematical literature, 
the Mabuchi action has been defined and used only for metrics having the same area or more generally, in higher dimensions, for metrics in the same K\"ahler class. The generalized definition \eqref{SMdef2} is natural in our case, because it is in this way that it will appear in the
calculations of the gravitational effective actions in Section 4. In particular, it satisfies the cocycle identities \eqref{cocM1} and \eqref{cocM2} for any metrics $g_{0}$ and $g$, including when $g_{0}$ and $g$ do not have the same areas.
It would be interesting to know if the higher dimensional versions of the action can also be suitably generalized as functions of the K\"ahler moduli consistently with the cocycle identities.

The variation of the Mabuchi action with respect to $g$ yields
\be\label{dM} \delta\SM  = 2\frac{\delta A}{A} - \ish (R-\bar R)\delta\phi\, ,\ee
where the constant $\bar R$ is the average Ricci scalar, or equivalently
the Ricci scalar for the metric of constant scalar curvature and area $A$, 
\be\label{Rbardef} \bar R = \frac{1}{A}\ish R = \frac{8\pi (1-h)}{A}\, \cdotp\ee
Comparing with \eqref{dSphi}, we see that the trace of the stress-energy
for the Mabuchi action satisfies
\be\label{tMcons} \Delta t_{\text M} = \frac{8\pi}{A}\,\bigl (R-\bar R
\bigr)\, ,\quad \ish t_{\text M}= 8\pi\, .\ee
The \emph{Ricci potential} $\psi$ is usually defined by the conditions
\begin{align}\label{Riccipotdef1} & \Delta \psi = R-\bar R\, ,\\&
\label{Riccipotdef2}\ish\psi= 0\, ,\end{align}
which always have a unique solution. Comparing \eqref{Riccipotdef1}, 
\eqref{Riccipotdef2} and \eqref{tMcons}, we find
\be\label{tMform} t_{\text M} = \frac{8\pi}{A}\bigl(\psi + 1)\, .\ee
\subsubsection{The Aubin-Yau action}

The Aubin-Yau action is defined by
\begin{align}\label{SAYdef1} \SAY (g_{0},\phi) &= -\ishbis\Bigl(\frac{1}{4}\phi\Delta_{0}\phi - \frac{\phi}{A_{0}}\Bigr)\\
\label{SAYdef2} & = -\frac{i}{2}\int_{\sigh}\partial\phi\wedge\bar\partial
\phi + \frac{1}{A_{0}}\int_{\sigh}\phi\omega_{0}\\
\label{SAYdef3} & = \frac{1}{2} \int_{\sigh}\phi\Bigl(
\frac{\omega_{0}}{A_{0}} + \frac{\omega}{A}\Bigr)\, .
\end{align}
As in the case of the Mabuchi action, we give here a definition
valid for metrics having different areas, slightly generalizing the functional used in the mathematical literature. 

It is important to realize that the Aubin-Yau action is not a functional of the metric $g$, because it is not invariant under 
constant shifts of $\phi$,
\be\label{shiftSAY} \SAY(g_{0},\phi + \text{constant}) = \SAY (g_{0},\phi) + 
\text{constant}\, .\ee
This is why we have been careful in \eqref{SAYdef1} to write $\SAY(g_{0},\phi)$ and not $\SAY(g_{0},g)$. In particular, the variation of $\SAY$ with respect to $\phi$,
\be\label{dAY} \delta\SAY  = \frac{1}{A}\ish \delta\phi\, ,\ee
is not of the form \eqref{dSphi}. 

However, 
$\SAY$ does satisfy cocycle identities, in the form
\be\label{cocAY} \SAY (g_{1},\phi_{1,3}) = \SAY(g_{1},\phi_{1,2}) 
+ \SAY(g_{2},\phi_{2,3})\, ,\ee
with $\omega _{2} = (A_{2}/A_{1})\omega_{1} + i A_{2}\partial\bar\partial
\phi_{1,2}$, $\omega _{3} = (A_{3}/A_{2})\omega_{2} + i A_{3}\partial\bar\partial \phi_{2,3}$ and $\phi_{1,3}=\phi_{1,2}+\phi_{2,3}$. Moreover, it is very easy to add simple terms to $\SAY$ to make it a well-defined functional of the metric. For example, a combination
\be\label{exAY} \SAY(g_{0},\phi) - \frac{1}{\hat A}\int_{\sigh}\!\sqrt{\hat g}\,\phi\, ,\ee
for any fixed metric $\hat g$ of area $\hat A$, is invariant under constant shifts of $\phi$ and satisfies the cocycle identities. The metric $\hat g$ could be for example the metric of constant curvature, or any other canonically defined metric on $\sigh$. As we shall see in the following, it is in the form \eqref{exAY} that the Aubin-Yau action contributes to the gravitational effective action of the model \eqref{dilatongen}.
Similarly, factors of the form
\be\label{gravdre} e^{K (\phi(x) - \SAY(g_{0},\phi))}\, ,\ee
for some constant $K$, are invariant 
under constant shifts of $\phi$ and satisfy the consistency conditions \eqref{cocycle3} and \eqref{cocycle4}. We shall demonstrate in Section 4 that these factors do contribute to the gravitational dressing of the operators \eqref{opalpha}, see eq.\ \eqref{dresslead}. 

\subsection{Variation formulas}
\label{psisec}

We are now going to present functionals $\Psi$ satisfying 
\eqref{solvecocycle} for the various actions introduced in the previous subsection, as well as some natural generalizations.

We denote by $\psi_{i}$, $i\geq 0$, the (real) eigenfunctions of the laplacian with eigenvalues $\lambda_{i}$, $0=\lambda_{0}<\lambda_{1}$, 
$\lambda_{i}\leq\lambda_{i+1}$. They are normalized such that
\be\label{eigenprod} \langle\psi_{i}|\psi_{j}\rangle = \ish\psi_{i}\psi_{j} = \delta_{ij}\, .\ee
In particular, the zero mode is
\be\label{zeromode}\psi_{0}=1/\sqrt{A}\, .\ee
The $\zeta$-function is defined by
\be\label{zetadef} \zeta(s) = \sum_{i>0}\frac{1}{\lambda_{i}^{s}}\,\cdotp\ee
The series in \eqref{zetadef} converges absolutely for $\re s >1$. By analytic continuation, it defines an analytic function $\zeta$ for all complex values of $s$, except at $s=1$ where it has a simple pole with residue $A/(4\pi)$.

\subsubsection{Partition function, determinant and the Liouville action}

As is well-known, the Liouville action can be written as
\be\label{SLPsi} \SL(g_{0},g) = \Psi_{\text{L}}(g) - 
\Psi_{\text{L}}(g_{0})
\ee
where the functional $\Psi_{\text{L}}$ is given by
\be\label{PsifromCFT} \Psi_{\text{L}}(g) = \frac{24\pi}{c}\ln Z_{\text{CFT}}(g)\ee
in terms of the partition function $Z_{\text{CFT}}$ of any conformal field theory of central charge $c$ in the background metric $g$. Equation \eqref{sgravCFT} is a special case of \eqref{PsifromCFT} with central charge  $c-26$ taking into account the contribution from the ghosts.

The simple case that will be useful for us corresponds to the $c=1$ CFT of a free massless scalar field, which yields
\be\label{PsiLdef} \Psi_{\text{L}}(g) = -12\pi\ln\frac{{\det}'\Delta_{g}}{A(g)}\, \cvp\ee
where $A(g)$ is the area and 
\be\label{detzeta} {\det}'\Delta_{g} = e^{-\zeta'(0)}\ee
is the infinite dimensional determinant of the laplacian with the zero mode excluded, defined in terms of the analytic continuation of the $\zeta$ function \eqref{zetadef}. Let us emphasize that the factor $1/A$ in \eqref{PsiLdef} is crucial for \eqref{SLPsi} to be valid.

The above results are standard and we shall not repeat the well-known proofs here.

\subsubsection{Green's functions and the Aubin-Yau action}

Let $G(x,y;g)$ be the Green's function for the laplacian $\Delta$ in the metric $g$. It is uniquely defined by the conditions
\begin{align}\label{green1} &\Delta_{x}G(x,y;g) = 
\frac{\delta(x-y)}{\sqrt{g}} - \frac{1}{A}\, \cvp\\\label{green2}
& \ish G(x,y;g) = 0
\end{align}
and can be expressed in terms of the eigenfunctions and eigenvalues of the laplacian as
\be\label{greendef} G(x,y;g) = \sum_{i>0}\frac{\psi_{i}(x)\psi_{i}(y)}{\lambda_{i}}\,\cdotp\ee
We are going to prove the following identity relating the variation of the Green's function to the Aubin-Yau action,
\be\label{PsiAY} G(x,y;g) - G(x,y;g_{0}) = \frac{1}{2}\Bigl(\phi(x) + \phi(y)\Bigr) - \SAY(g_{0},\phi)\, .\ee
Let us note that the left-hand side of the above equation is a well-defined functional of the metric $g$ and thus the right-hand side must
be invariant under constant shifts in $\phi$, which is indeed the case from the transformation rule \eqref{shiftSAY}.

There are many ways to prove \eqref{PsiAY}. The simplest is to check that the conditions \eqref{green1} and \eqref{green2} are satisfied for 
$G(x,y;g)$ given by \eqref{PsiAY}, if they are 
satisfied for $G(x,y;g_{0})$. To check \eqref{green1} we use 
\eqref{laprel} and \eqref{phidef},
\begin{align} \Delta_{x}G(x,y;g) &= e^{-2\sigma}\Bigl(\Delta_{0x}G(x,y;g_{0}) + \frac{1}{2}\Delta_{0}\phi(x)\Bigr)\\
& = e^{-2\sigma}\Bigl( \frac{\delta(x-y)}{\sqrt{g_{0}}} - \frac{1}{A_{0}}
+\frac{1}{A_{0}} - \frac{e^{2\sigma}}{A}\Bigr)\\
& = \frac{\delta(x-y)}{\sqrt{g}} - \frac{1}{A}\,\cdotp
\end{align}
To check \eqref{green2}, we use $\sqrt{g} = e^{2\sigma}\sqrt{g_{0}}$ and
\eqref{phidef} to write
\begin{multline}\ish G(x,y;g) = \int_{\sigh}\!\d^{2}x\,\sqrt{g_{0}}
\Bigl(\frac{A}{A_{0}} - \frac{1}{2}A\Delta_{0}\phi(x)\Bigr)
\Bigl( G(x,y;g_{0}) + \frac{1}{2}\phi(x)\Bigr)
\\+\frac{A}{2}\phi(y) - A\SAY(g_{0},\phi)\, .
\end{multline}
We then integrate by part the term containing $\Delta_{0}\phi$, use the conditions \eqref{green1} and \eqref{green2} for $g=g_{0}$ and the definition of the Aubin-Yau action \eqref{SAYdef1} to get \eqref{green2} 
for $g$.

An alternative derivation would be to prove the infinitesimal version of \eqref{PsiAY},
\be\label{psiAYinf} \delta G(x,y;g) = \frac{1}{2}\Bigl(\delta\phi(x) 
+ \delta\phi(y)\Bigr) - \ish\delta\phi\, .\ee
This can be done starting from the defining equation \eqref{greendef} and using the quantum mechanical perturbation theory formulas
for the infinitesimal variations of the eigenvalues and eigenfunctions of the laplacian
\begin{align}\label{pertQM1}\delta\lambda_{i}&= -2\lambda_{i}
\langle\psi_{i}|\delta\sigma|\psi_{i}\rangle\, ,\\\label{pertQM2}
\delta\psi_{i} & =-\langle\psi_{i}|\delta\sigma|\psi_{i}\rangle
\psi_{i} - 2\sum_{j\not = i}\frac{\lambda_{i}}{\lambda_{i}-\lambda_{j}}
\langle\psi_{j}|\delta\sigma|\psi_{i}\rangle \psi_{j}\, .
\end{align}
Deriving \eqref{psiAYinf} is then a straightforward calculation that we let to the reader. Let us note that the formulas \eqref{pertQM1} and \eqref{pertQM2} are valid only when the spectrum is non-degenerate, but this is of course almost always true and \eqref{PsiAY} follows in all cases by continuity.

When $x\rightarrow y$, the Green's function has the usual short distance logarithmic divergence. This divergence can be subtracted in a diffeomorphism invariant way, by using the geodesic distance function
$d_{g}$ and introducing an arbitrary, metric-independent, length scale
$\ell$. 
The renormalized Green's function at coincident points $G_{R}(x)$ is then
defined by
\be\label{GRdef1} G_{R}(x;g) = \lim_{y\rightarrow x}\Bigl[
G(x,y;g) + \frac{1}{2\pi}\ln\frac{d_{g}(x,y)}{\ell}\Bigr]\,\cdotp\ee
We shall also use later an essentially equivalent definition in terms 
of $\zeta$ functions, see \eqref{GRzetadef} and \eqref{GRGRrel}.
This formula together with \eqref{PsiAY} immediately yield
\be\label{PsiAY2} G_{R}(x;g) - G_{R}(x,g_{0}) = \phi(x) - \SAY (g_{0},\phi) + \frac{\sigma(x)}{2\pi}\,\cvp\ee
the last term $\sigma/(2\pi)$ coming from the variation of the geodesic distance under rescaling of the metric.

The formulas \eqref{PsiAY} and \eqref{PsiAY2} could also be derived starting from standard general formulas for the Weyl rescaling of the Green's function that can be found for example in \cite{reviewstrings}.

\subsubsection{The Ricci potential and the Mabuchi and Aubin-Yau actions}

The Ricci potential associated with a given metric was defined by the conditions \eqref{Riccipotdef1} and \eqref{Riccipotdef2}.
The difference between Ricci potentials associated with different metrics is given by an interesting formula involving both the Aubin-Yau and Mabuchi actions,
\be\label{psiricdiff} \psi(x;g) - \psi(x;g_{0}) = 2\sigma (x) + 4\pi (1-h)
\bigl(\phi (x) - \SAY (g_{0},\phi)\bigr) - \frac{1}{2}\SM(g_{0},g)\, .\ee
To prove this identity, we first check that 
\be \Delta\bigl(\psi(x;g_{0}) + 2\sigma (x) + 4\pi (1-h)
\phi (x)\bigr) = R-\bar R\ee
using $\Delta_{0}\psi(x;g_{0}) = R_{0}-\bar R_{0}$, \eqref{Rrelconf} and \eqref{phidef}. This implies that
\be\label{psiricderive} \psi(x;g) - \psi(x;g_{0}) = 2\sigma (x) + 4\pi (1-h)
\phi (x) + C(g_{0},g)\, ,\ee
for some $x$-independent functional $C(g_{0},g)$. This functional is then obtained by computing the integral of \eqref{psiricderive} and imposing the condition \eqref{Riccipotdef2} for both $\psi(x;g)$ and $\psi(x;g_{0})$, using in particular \eqref{phidef}.

Let us note that if $\bar g$ is the metric of constant scalar curvature, $\psi(x;\bar g) = 0$. Equation \eqref{psiricdiff} then yields the Ricci
potential for any metric $g=e^{2\sigma}\bar g$,
\be\label{Riccpfor} \psi(x;g) = 
2\sigma (x) + 4\pi (1-h)
\bigl(\phi (x) - \SAY (\bar g,\phi)\bigr) - \frac{1}{2}\SM(\bar g,g)\, .\ee

Let us also note that the Ricci potential is given by the following explicit integral formula
\be\label{ricciintf} \psi(x;g) = \int_{\sigh}\!\d y\sqrt{g(y)}\, G(x,y;g) R(y)\ee
in terms of the Green's function. Equation \eqref{psiricdiff} could also be straightforwardly derived from this integral representation by using
\eqref{PsiAY}.

\subsubsection{The Polyakov functional and the Mabuchi action}

The next functional we wish to study is Polyakov's effective action
\be\label{poldef} \Psi_{\text{P}}(g) = \frac{1}{4}\int_{\sigh\times\sigh}\!\d^{2}x\d^{2}y\sqrt{g(x)}\sqrt{g(y)}\, R(x)G(x,y;g)R(y)\, .\ee
This is the famous non-local action
$\iint R\Delta^{-1}R$ introduced by Polyakov in \cite{polyakov} to describe the partition function of a CFT. However, when the theory is formulated at finite area,
it turns out that the variation of $\Psi_{\text{P}}$ crucially involves the Mabuchi action on top of the Liouville action,
\be\label{PsiM} \Psi_{\text{P}}(g) - \Psi_{\text{P}}(g_{0})=
\SL(g_{0},g) -2\pi(1-h) \SM(g_{0},g)\, .\ee
In particular, $\Psi_{\text P}$ as defined in \eqref{poldef} is \emph{not} suitable to describe the partition function of a CFT, whose variation 
\eqref{SLPsi} always yields the Liouville action without any additional term. When $A\rightarrow\infty$, the Mabuchi term vanishes because in this limit \eqref{phidef} shows that $\phi$ scales like $1/A$.

The derivation of \eqref{PsiM} is a bit tedious but completely straightforward. One starts by expressing $\Psi_{\text P}(g)$ in terms of quantities related to $g_{0}$, by using equations \eqref{Rrelconf}, \eqref{PsiAY}, \eqref{green1}, \eqref{green2} and \eqref{phidef}. Terms involving $\Delta_{0}\sigma$ or $\Delta_{0}\phi$ are dealt with by integrating by part and using repeatedly \eqref{green1}. The genus $h$ enters from integrals of the form \eqref{euler}.

A very simple expression for the functional \eqref{poldef} can also be obtained in terms of the Ricci potential $\psi$ defined in \eqref{Riccipotdef1} and \eqref{Riccipotdef2}, by replacing $R$ by 
$\bar R + \Delta\psi$ in \eqref{poldef},
\be\label{PolfRic} \Psi_{\text P}(g) = \frac{1}{4}\ish
\psi\Delta\psi \, .\ee
This formula, together with \eqref{psiricdiff}, is another good starting point to derive \eqref{PsiM}.

\subsubsection{The integrated Green's function}

The last non-trivial functional that we shall need in Section 4 is the integral of the Green's function at coincident points,
\be\label{psinew}\Psi_{\text G}(g) = \frac{1}{A(g)}\ish G_{R}(x;g)\, .\ee
By using \eqref{phidef} and \eqref{PsiAY2}, as well as the explicit formula \eqref{SAYdef1} for $\SAY$, we easily obtain
\be\label{psiG1} \Psi_{\text G}(g) - \Psi_{\text G}(g_{0}) =
\frac{1}{8\pi}\int_{\sigh}\!\d^{2}x\sqrt{g_{0}}\,\biggl[ -2\pi\phi\Delta_{0}\phi
+\frac{4}{A}\sigma e^{2\sigma} - 4\pi\phi\Delta_{0}G_{R}(x;g_{0})\biggr]\, .\ee
This formula has also been derived and used in \cite{M} using different methods.

For our purposes, we need to go further and evaluate $\Delta G_{R}$. We present first the cases of the sphere and the torus, for which a very elementary calculation is available.

\noindent\emph{The case of the sphere}

On the sphere endowed with the round metric $\bar g$, the $\text{SO}(3)$ invariance implies that the scalar function $G_{R}(x;\bar g)$ must be a constant. Applying the Laplace operator on the equation \eqref{PsiAY2} with $g_{0}=\bar g$ thus yields 
\be\label{spreas}\Delta G_{R}(x;g) = \Delta\phi + \frac{1}{2\pi}\Delta\sigma =
\frac{1}{4\pi}\Delta\psi\, ,\ee
where the Ricci potential on the sphere is given by \eqref{Riccpfor} for $h=0$. Using \eqref{Riccipotdef1} we thus conclude that
\be\label{DeltaGRsphere} \Delta G_{R} = \frac{R-\bar R}{4\pi} =\frac{R}{4\pi} - \frac{2}{A}\quad \text{on the sphere}\, .\ee

Plugging this result for $g=g_{0}$ into \eqref{psiG1} and comparing the result with the definition \eqref{Mabdef1} of the $h=0$ Mabuchi action then yields
\be\label{psiGd1} \Psi_{\text G}(g) - \Psi_{\text G}(g_{0}) =
\frac{1}{8\pi}\SM(g_{0},g)\, .\ee

\noindent\emph{The case of the torus}

On the torus and for $g_{0}=\bar g$ the flat metric, $G_{R}(x;\bar g)$
is a constant by translation invariance. The first equation in \eqref{spreas} is thus still valid, but the formula \eqref{Riccpfor} for the Ricci potential in the case $h=1$ yields in the present case
\be\label{lapGRt1} \Delta G_{R}(x;g) = \frac{1}{4\pi}\Delta\psi + 
\Delta\phi = \frac{R}{4\pi} - \frac{2}{A} + \frac{2}{\bar A} e^{-2\sigma}\, ,\ee
where $\bar A$ is the area for the flat metric $\bar g$ and we have used 
\eqref{phidef} and \eqref{laprel}. Let us introduce the usual coordinates
$(z,\bar z)$ on the torus, with the identifications $z\equiv z+1$ and $z\equiv z+\tau$ for a complex structure modulus $\tau$ such that $\im\tau >0$. The flat metric of area $\bar A$ is simply
\be\label{flatmettor} \bar g = \frac{\bar A}{\im\tau} |\d z|^{2}\ee
and thus \eqref{lapGRt1} can be rewritten in the nice form
\be\label{lapGRt2} \Delta G_{R}(x;g) =
\frac{R}{4\pi} - \frac{2}{A} + \frac{2}{\sqrt{g}} \frac{1}{\im\tau}=
\frac{R}{4\pi} - \frac{2}{A} + 2\frac{\sqrt{g_{\text c}}}{\sqrt{g}} 
\,\cvp\ee
where
\be\label{gctorus} g^{\text c} = \frac{1}{\im\tau} |\d z|^{2}\ee
is the so-called \emph{canonical} metric on the torus, the flat metric
normalized to have unit area.

Plugging \eqref{lapGRt2} into \eqref{psiG1} and comparing with the definitions of the $h=1$ Mabuchi \eqref{Mabdef1} and Aubin-Yau 
\eqref{SAYdef1} actions, we obtain
\be\label{psiGd2} \Psi_{\text G}(g) - \Psi_{\text G}(g_{0}) =
\frac{1}{8\pi}\SM(g_{0},g) + \SAY(g_{0},\phi) - \int_{\sigh}\!\d^{2}x
\sqrt{g^{\text c}}\,\phi\, .\ee
The Aubin-Yau term in the above formula is of the form \eqref{exAY}.

\noindent\emph{The case of a general compact Riemann surface}

For an arbitrary genus $h$, \eqref{psiGd1} and \eqref{psiGd2} generalize to
\be\label{psiGdall} \Psi_{\text G}(g) - \Psi_{\text G}(g_{0}) =
\frac{1}{8\pi}\SM(g_{0},g) + h\Bigl(\SAY(g_{0},\phi) - \int_{\sigh}\!\d^{2}x\sqrt{g^{\text c}}\,\phi\Bigr)\, .\ee

The \emph{canonical} metric $g^{\text c}$ on $\sigh$ is defined as follows. We introduce a canonical basis of homology cycles $(\alpha_{i},\beta_{i})$, $1\leq i\leq h$, on $\sigh$, with non-trivial intersection numbers 
$\langle\alpha_{i},\beta_{j}\rangle = \delta_{ij}$. We also introduce a basis of holomorphic one-forms $\lambda_{i}$, $1\leq i\leq h$, such that
\be\label{onefnorm} \oint_{\alpha_{i}}\lambda_{j} = \delta_{ij}\, .\ee
The K\"ahler form of the canonical metric is then defined to be
\be\label{Kahlercan} \omega^{\text c}= \frac{i}{2h} (\im\tau)^{-1}_{ij}
\lambda_{i}\wedge\bar\lambda_{j}\ee
in terms of the period matrix
\be\label{period} \tau_{ij} = \oint_{\beta_{i}}\lambda_{j}\, .\ee
The formula \eqref{Kahlercan} indeed defines a metric because the matrix $\tau$ is symmetic and $\im\tau$ is positive-definite; in a coordinate system $(z,\bar z)$ in which
\be\label{lambcoor}\lambda_{i} = l_{i}(z)\d z\, ,\ee
we have
\be\label{gczz} 2g^{\text{c}}_{z\bar z} =\sqrt{g^{\text c}}= \frac{1}{h} (\im\tau)^{-1}_{ij}
l_{i}(z)\bar l_{j}(\bar z) >0\, .\ee
Let us note that the definition \eqref{Kahlercan} is of course independent of the choice of basis of holomorphic one-forms satisfying \eqref{onefnorm}, and that the overall normalization is chosen in such a way that $g^{\text c}$ has unit area, as can be easily shown using Riemann bilinear relations. The canonical metric can also be interpreted as being the pull-back of the trivial metric on the Jacobian variety of $\sigh$ under the Abel-Jacobi map.

Using \eqref{psiG1}, equation \eqref{psiGdall} follows from the following general formula for the laplacian of $G_{R}$,
\be\label{DGRgen} \Delta G_{R}(x;g) = \frac{R}{4\pi} - \frac{2}{A}
+ 2h\frac{\sqrt{g^{\text c}}}{\sqrt{g}}\,\cdotp\ee

If we try to repeat the reasoning made at genus zero and one to prove this formula, applying the Laplace operator on \eqref{PsiAY2} and using \eqref{phidef} and \eqref{laprel}, we get a nice transformation rule
\be\label{Grtrans} \Delta G_{R}(x;g) - \frac{R}{4\pi} + \frac{2}{A} =
e^{-2\sigma}\Bigl( \Delta G_{R}(x;g_{0}) - \frac{R_{0}}{4\pi} + \frac{2}{A_{0}}\Bigr)\, .\ee
However, this result would yield $\Delta G_{R}$ for any metric if it were known for a particular metric on $\sigh$, which is not the case for $h\geq 2$. We thus have to proceed in a different way.

Let us consider the function
\be\label{GRsfdef}\mathsf G_{R}(x,y;g) = G(x,y;g) + \frac{1}{2\pi}
\ln\frac{d_{g}(x,y)}{\ell}\,\cdotp\ee
It is symmetric, $\mathsf G_{R}(x,y;g)=\mathsf G_{R}(y,x;g)$, and
regular at $x=y$. The definition \eqref{GRdef1} is equivalent to
\be\label{GRsfGR} G_{R}(x;g) = \mathsf G_{R}(x,x;g)\, .\ee
Let us pick a coordinate system $(x^{a})$ for which, in some open set, the metric has the simple form
\be\label{metscoo} g_{ab}=e^{2\sigma}\delta_{ab}\ee
and thus the laplacian reads
\be\label{lapspecial}\Delta = -e^{2\sigma}\partial_{a}\partial_{a}\, .\ee
Taking the partial derivatives of \eqref{GRsfGR} and using the symmetry of $\mathsf G_{R}$ we get
\be\label{partialGR} \partial_{a}\partial_{b}G_{R}(x;g) =
2(\partial_{1a}\partial_{1b} + \partial_{1a}\partial_{2b})
\mathsf G_{R}(x,x;g)\, ,\ee
where $\partial_{1a}$ and $\partial_{2a}$ are partial derivatives with 
respect to the first and second argument in $\mathsf G_{R}(x,y;g)$.
The implies in particular that
\begin{multline}\label{DgRa} \Delta G_{R}(x;g) = 2\lim_{y\rightarrow x}
\biggl[
\Delta_{1}G(x,y;g) + \frac{1}{2\pi}\Delta_{1}\ln d_{g}(x,y)\\
- e^{-2\sigma}\partial_{1a}\partial_{2a}G(x,y;g) - \frac{1}{2\pi}
e^{-2\sigma}\partial_{1a}\partial_{2a}\ln d_{g}(x,y)\biggr]\, ,
\end{multline}
where $\Delta_{1}$ acts on the first argument $x$.
The terms involving the geodesic distance can be evaluated using the standard short distance expansion in the coordinate system \eqref{metscoo},
\begin{multline}\label{geodexp} 2\ln d_{g}(x,y) = \ln \bigl((y^{a}-x^{a})(y^{a}-x^{a})
\bigr) +
2\sigma(x) + (y^{a}-x^{a})\partial_{a}\sigma(x)\\ - \frac{1}{12}
(y^{a}-x^{a})(y^{a}-x^{a})\partial_{b}\sigma(x)\partial_{b}\sigma(x) 
+ \frac{1}{6}
(y^{a}-x^{a})(y^{b}-x^{b})\partial_{a}\sigma(x)\partial_{b}\sigma(x)\\
+\frac{1}{3}(y^{a}-x^{a})(y^{b}-x^{b})\partial_{a}\partial_{b}\sigma(x)
+ \mathcal O\bigl((y-x)^{3}\bigr)\, ,
\end{multline}
which yields
\begin{align}\label{Dgeod1} \frac{1}{2\pi}\Delta_{1}\ln d_{g}(x,y) &
\underset{x\rightarrow y}{=}
-\frac{\delta(x-y)}{\sqrt{g}} + \frac{R}{12\pi} \, \cvp\\
\label{Dgeod2} \frac{1}{2\pi}e^{-2\sigma}\partial_{1a}\partial_{2a}\ln
d_{g}(x,y)& \underset{x\rightarrow y}{=} 
-\frac{\delta(x-y)}{\sqrt{g}} - \frac{R}{24\pi} \, \cdotp
\end{align}
Using \eqref{green1}, we thus obtain
\be\label{DeltaGR} \Delta G_{R}(x;g) = 2\lim_{x\rightarrow y}\biggl[
\frac{\delta(x-y)}{\sqrt{g}} + \frac{R}{8\pi} - \frac{1}{A}
-e^{-2\sigma}\partial_{1a}\partial_{2a} G(x,y;g)\biggr]\, .\ee
Equation \eqref{DGRgen} then follows from the identity
\be\label{mixedpart} \partial_{1a}\partial_{2a}G(x,y;g) 
\underset{x\rightarrow y}{=}  \delta(x-y) -
(\im\tau)^{-1}_{ij}l_{i}\bar l_{j}\, .\ee
This identity is a direct consequence of a
standard formula \eqref{pderG} for the mixed partial derivatives of the Green's function in a coordinate system \eqref{metscoo}.
This formula can be found for example in \cite{reviewstrings}. Since we have not been able to find a simple derivation in the literature, we have included one in Appendix A.

\section{Application to the massive scalar field}
\label{massivescalarsec}

We are now going to apply the results of the previous Section to study the model \eqref{dilatongen}. In order to compute both the effective gravitational action and the gravitational dressing of the operators 
\eqref{opalpha} at the same time, we consider the generalized partition function
\be\label{Zdil}\Zmat(q,m; \k,\x;g) = \int\!\mathscr D X\, e^{\sum_{i=1}^{n}k_{i}X(x_{i}) - \Smat(X;q,m;g)}\, ,\ee
where $\k$ and $\x$ denote collectively all the $k_{i}$s and $x_{i}$s.
This generalized partition function contains all the information we need.
Indeed, the usual matter partition function is simply
\be\label{Zmatdil} \Zmat(q,m;g) = \Zmat(q,m;\k=0,\x;g)\, .\ee
Taking into account the contribution from the ghost CFT, the gravitational effective action \eqref{Sgravdef} is given by
\be\label{Sgravlindil} \Sgrav(g_{0},g;q,m) = \frac{13}{12\pi}\SL(g_{0},g) -\ln\frac{\Zmat(q,m;g)}{\Zmat(q,m;g_{0})}\,\cdotp\ee
The correlators for the operators \eqref{opalpha}, defined in \eqref{matcorrdef}, are given by
\be\label{corrlindil}
\bigl\langle \mathscr O_{k_{1}}(x_{1})\cdots
\mathscr O_{k_{n}}(x_{n})\bigr\rangle_{\text{mat},\, g} = 
\frac{\Zmat(q,m; \k,\x;g)}{\Zmat(q,m;g)}\,\cdotp\ee
Finally, the gravitational dressing \eqref{dressingdef} is
\begin{multline}\label{dressinglindil}
D_{\mathscr O_{k_{1}}\cdots\mathscr O_{k_{n}}}(\x,\k;
q,m; g_{0},g) =\\ 
\frac{\Zmat(q,m; \k,\x;g)}{\Zmat(q,m; \k,\x;g_{0})}
\, e^{2\sum_{i=1}^{n}\sigma(x_{i}) -\frac{13}{12\pi}
\SL(g_{0},g) + \Sgrav(g_{0},g;q,m)}\, .\end{multline}

We are going to study the model \emph{at leading non-trivial order when the mass parameter is small.} If we work at fixed area, as in \eqref{pi2}, this means that the condition
\be\label{smallm1} m^{2}A \ll 1\ee
must be satisfied. If we wish to integrate over areas, then we choose the cosmological constant to be much larger than $m^{2}$,
\be\label{smallm2} m^{2}\ll\mu\, .\ee
It is important to understand that this mass expansion is non-perturbative. Indeed, the mass term in \eqref{dilatongen} is not a well-defined operator in the $m=0$ CFT.

\subsection{Direct calculation}
\label{directsec}

\subsubsection{The partition function}

We introduce the Green's function $G(x,y;m;g)$ for the massive scalar field, characterized by the condition
\be\label{greenm1} (\Delta_{x}+m^{2}) G(x,y;m;g) = 
\frac{\delta(x-y)}{\sqrt{g}}\ee
and expressed explicitly in terms of the eigenfunctions and eigenvalues of the laplacian as
\be\label{greenmdef} G(x,y;m;g) = \sum_{i\geq0}\frac{\psi_{i}(x)\psi_{i}(y)}{\lambda_{i}+m^{2}}\,\cdotp\ee
The path integral \eqref{Zdil} is gaussian, so a direct calculation of the partition function in terms of the above Green's function is completely straightforward. The only subtlety is that we have to subtract the infinities due to the self-contractions. We proceed as usual (``normal ordering''), in a diffeomorphism invariant way, by replacing each instance of $G(x,x;m;g)$ by its renormalized version $G_{R}(x;m;g)$, defined along the lines of
\eqref{GRdef1},
\be\label{GRdefmass} G_{R}(x;m;g) = \lim_{y\rightarrow x}\Bigl[
G(x,y;m;g) + \frac{1}{2\pi}\ln\frac{d_{g}(x,y)}{\ell}\Bigr]\,\cdotp\ee
The parameter $\ell$ plays the r\^ole of an arbitrary renormalization scale. We also define the renormalized determinant of the euclidean Klein-Gordon operator with the $\zeta$-function prescription, along the lines of \eqref{zetadef} and \eqref{detzeta},
\be\label{detmzeta} \det (\Delta + m^{2}) = e^{-\zeta_{m}'(0)}\, ,\ee
with
\be\label{zetamdef} \zeta_{m}(s) = \sum_{i\geq 0}\frac{1}{(\lambda_{i}+m^{2})^{s}}\,\cdotp\ee
The partition function \eqref{Zdil} then reads
\begin{multline}\label{Zdilexact}\Zmat(q,m; \k,\x;g) =
\frac{1}{\sqrt{\det(\Delta + m^{2})}}\\
\exp\biggl[2\pi\sum_{i\not = j}k_{i}k_{j}G(x_{i},x_{j};m;g) + 
2\pi\sum_{i}k_{i}^{2}G_{R}(x_{i};m;g) -\\
\frac{q}{2}\sum_{i}k_{i}\ish G(x,x_{i};m;g)R(x)
+\\\frac{q^{2}}{32\pi}\int_{\sigh\times\sigh}\d^{2}x\d^{2}y\sqrt{g(x)}
\sqrt{g(y)}\, R(x)G(x,y;m;g)R(y)\biggr]\, .
\end{multline}

\subsubsection{The small mass expansion}

To pick up the leading non-trivial contributions at small $m$, we need to approximate the Green's functions \eqref{greenmdef}, 
\eqref{GRdefmass} and the determinant \eqref{detmzeta}. The small mass expansion of the Green's functions is straightforward to obtain. Taking care of the zero mode \eqref{zeromode} and using \eqref{greendef} and \eqref{GRdef1} we get
\begin{align}
\label{smallmG} &G(x,y;m;g) = \frac{1}{m^{2}A} + G(x,y;g) + 
m^{2}G_{2}(x,y;g) + \mathcal O(m^{4})\, ,\\
\label{smallmGR}
&G_{R}(x;m;g) = \frac{1}{m^{2}A} + G_{R}(x;g) + 
m^{2}G_{2}(x,x;g) + \mathcal O(m^{4})\, ,
\end{align}
where $G_{2}(x,y;g)$ is defined by
\be\label{G2def} G_{2}(x,y;g) = -\sum_{i>0}\frac{\psi_{i}(x)\psi_{i}(y)}{\lambda_{i}^{2}}\,\cdotp\ee
On the other hand, from \eqref{detmzeta} and the small mass expansion of the $\zeta$-function \eqref{zetamdef}, we get
\be\label{detmexp} \ln\det (\Delta + m^{2}) = \ln(m^{2}{\det}'\Delta) + 
m^{2}\lim_{s\rightarrow 0}\Bigl[ \zeta(s+1) + s\zeta'(s+1)\Bigr]
+ \mathcal O\bigl( m^{4}\bigr)\, .\ee
The second term on the right hand side gives a renormalized version of the formal sum $\sum_{i>0}1/\lambda_{i}$ or equivalently of the integrated Green's function $\int\!\d^{2}x\sqrt{g}\,G(x,x;g)$. Another renormalized version of the same functional is given by $A\Psi_{\text G}(g)$ defined in \eqref{psinew}. The difference between these two definitions of $\sum_{i>0}1/\lambda_{i}$, based on two different renormalization schemes (explicit subtraction of the UV divergence in the definition \eqref{psinew} using $G_{R}$ \eqref{GRdef1} or $\zeta$-function in \eqref{detmexp}) must be a finite local counterterm, which presently can only correspond to a finite shift of the cosmological constant. Indeed, we show explicitly in Appendix B that
\be\label{zetadschemes}
\lim_{s\rightarrow 0}\Bigl[ \zeta(s+1) + s\zeta'(s+1)\Bigr] = A
\Bigl(\Psi_{\text G}(g) +\frac{1}{2\pi}\bigl(\ln\ell-\ln 2 + \gamma\bigr)
\Bigr)\, , \ee
where $\gamma \simeq 0.577$ is Euler's constant.

\subsubsection{The gravitational action}

Plugging the expansions \eqref{smallmG} and \eqref{detmexp} into \eqref{Zdilexact} for $\k=0$ we obtain
\begin{multline}\label{lnZmat} -\ln\Zmat (q,m;g) = -\frac{2\pi (1-h)^{2}q^{2}}{m^{2}A} + \frac{1}{2}\ln\bigl( m^{2}{\det}'\Delta\bigr)
-\frac{q^{2}}{32\pi}\Psi_{\text P}(g)\\ +
\frac{1}{2} Am^{2}
\Bigl(\Psi_{\text G}(g) +\frac{1}{2\pi}\bigl(\ln\ell-\ln 2 + \gamma\bigr)
\Bigr)\\ - \frac{q^{2}m^{2}}{32\pi}
\int_{\sigh\times\sigh}\!\d^{2}x\d^{2}y\sqrt{g(x)}\sqrt{g(y)}\, R(x)
G_{2}(x,y;g)R(y) +
\mathcal O\bigl(m^{4}\bigr)\, .\end{multline}

\paragraph{Case $q\not = 0$}

The \emph{leading} non-trivial contributions are then given by the first line in \eqref{lnZmat}. From \eqref{SLPsi}, \eqref{PsiLdef} and \eqref{PsiM}, we obtain the leading gravitational action
\begin{multline}\label{sleadq} \Sgrav(g_{0},g;q,m) =
-\frac{2\pi(1-h)^{2}q^{2}}{m^{2}}\Bigl(\frac{1}{A}-\frac{1}{A_{0}}\Bigr)+\frac{1}{2}\ln\frac{A}{A_{0}} \\ +\frac{25-3q^{2}}{24\pi}\SL(g_{0},g) 
+\frac{(1-h)q^{2}}{4}\SM(g_{0},g)\, .
\end{multline}

The terms in $1/A$ in the action come from the integration over the zero mode of the scalar field $X$. These terms are seen in $R^{2}$ gravity models 
\cite{R2grav}, because an $R^{2}$ term in the bare gravitational action
is equivalent to the model \eqref{dilatongen} with no kinetic term and an imaginary $q$. For real $q$, the effect of these
terms in to enhance the contribution of surfaces of small areas, consistently with our approximation \eqref{smallm1}. In the $m\rightarrow 0$ CFT limit, the integral over the zero mode imposes the ``neutrality'' condition 
$q=0$.

The coefficient $3q^{2}-25 = 1+3q^{2}-26$ in front of the Liouville action is the total central charge of the conformal $m=0$ model coupled with the ghost CFT. This coefficient 
can be made to vanish, either by adjusting $q$, or by coupling the system
to a spectator CFT of central charge $c$ such that
\be\label{Mabdom} 1+3q^{2}+c-26 = 0\, .\ee
For $h\not = 1$, the non-trivial part of the gravitational action is then entirely given by the Mabuchi action. If $q$ is real and $h=0$, the Mabuchi action comes with the right sign to define non-perturbatively a new
two-dimensional quantum gravity model, because $\SM$ is bounded from below and convex on the space of metrics. When $h\geq 2$, $q$ must be imaginary.
In this case positivity is lost and we must work at fixed area for the small mass approximation to make sense, because of the $1/A$ term in the action. If $h\geq 2$ and $q$ is real, higher order terms in the small mass expansion cannot be neglected non-perturbatively, since otherwise the gravitational action wouldn't be bounded from below.

\paragraph{Case $q = 0$} The leading non-trivial contribution to the
gravitational action is then given by the second line in \eqref{lnZmat}.
From \eqref{psiGdall} we get
\begin{multline}\label{sleadm} \Sgrav(g_{0},g;m) = 
\frac{1}{2}\ln\frac{A}{A_{0}} + 
\frac{m^{2}}{4\pi}\bigl(\ln\ell -\ln 2 + \gamma 
+ 2\pi\Psi_{\text G}(g_{0})\bigr)
\bigl(A-A_{0}\bigr)\\ + \frac{25}{24\pi}\SL(g_{0},g) 
+\frac{m^{2}A}{16\pi}\biggl[\SM(g_{0},g) + 8\pi h\Bigl(
\SAY(g_{0},\phi) - \int_{\sigh}\!\d^{2}x\,\sqrt{g^{\text c}}\,\phi\Bigr)
\biggr]\, .
\end{multline}
As in the previous case, we may cancel the Liouville term in the gravitational action by coupling to a spectator CFT. The non-trivial part of the gravitational action is then entirely dominated by the Mabuchi action in the case of a spherical space-time or by its generalized version including the Aubin-Yau and canonical metric terms in higher genuses.

\paragraph{General case}

If we want to keep all terms of orders $m^{2}$ when $q\not = 0$, we have to deal with the functional
\be\label{Psilast}
\Psi(g)=
\int_{\sigh\times\sigh}\!\d^{2}x\d^{2}y\sqrt{g(x)}\sqrt{g(y)}\, R(x)
G_{2}(x,y;g)R(y)\ee
which appears in the last line of \eqref{lnZmat}. Using the conditions
\begin{align}\label{glevel2} &\Delta_{x}G_{2}(x,y;g) = 
-G(x,y;g)\, \cvp\\\label{glevel22}
& \ish G_{2}(x,y;g) = 0\, ,
\end{align}
which follow readily from the definition \eqref{G2def}, it is straightforward to express $\Psi$ in terms of the Ricci potential,
\be\label{pvsRic} \Psi(g) = -\ish \psi^{2}\, .
\ee
The variation of $\Psi(g)$ and in particular its contribution
\be\label{psilevcont} -\frac{q^{2}m^{2}}{32\pi}\bigl(\Psi(g)-\Psi(g_{0})
\bigr)\ee
to the gravitational action can then be computed straightforwardly from
\eqref{psiricdiff}, but the resulting formula is not particularly illuminating. The most elegant form is obtained when $g_{0}=\bar g$ is 
the metric of constant curvature, in which case \eqref{Riccpfor} shows that \eqref{psilevcont} yields the contribution
\be\label{sleadmq} \frac{q^{2}m^{2}}{32\pi}\ish\Bigl(
2\sigma + 4\pi (1-h)\bigl(\phi - \SAY(\bar g,\phi)\bigr) - \frac{1}{2}\SM(\bar g,g)
\Bigr)^{2}\ee
to the gravitational effective action.

\subsubsection{The gravitational dressing}

Let us introduce
\be\label{aqdef}\alpha_{q,\k}= \sum_{i}k_{i}- (1-h)q\, .\ee
Let us note that in the CFT case, i.e.\ in the absence of mass term, the integration over the scalar field zero
mode imposes the condition
\be\label{CFTcond} \alpha_{q,\k}=0\, ,\ee
whereas when $m\not = 0$ this parameter can be arbitrary.

The $\k$-dependent terms in \eqref{Zdilexact}, in the small mass expansion, read
\begin{multline}\label{kdepsm} \ln\Zmat (q,m;\k,\x;g) =
-\frac{2\pi (1-h)^{2} q^{2}}{m^{2}A} + \frac{2\pi\alpha_{q,\k}^{2}}{m^{2}A}\\
+ 2\pi\sum_{i\not = j}k_{i}k_{j}G(x_{i},x_{j};g) + 2\pi\sum_{i}k_{i}^{2}
G_{R}(x_{i};g)- \frac{q}{2}\sum_{i}k_{i}\psi(x_{i};g)\\
+ 2\pi m^{2}\sum_{i,j}k_{i}k_{j}G_{2}(x_{i},x_{j};g) - \frac{q}{2} m^{2}
\sum_{i}k_{i}\ish G_{2}(x_{i},x;g) R(x)\\
+\mathcal O\bigl(m^{4}\bigr) + \k\text{-independent terms.}
\end{multline}

The leading non-trivial contributions
to the gravitational dressing factor \eqref{dressinglindil} are
given by the second line in \eqref{kdepsm}. From \eqref{PsiAY}, 
\eqref{PsiAY2} and \eqref{psiricdiff} we get
\begin{multline}\label{dresslead}
\ln D_{\mathscr O_{1}\cdots\mathscr O_{n}}(\x,\k,q; g_{0},g)=
-\frac{2\pi (1-h)^{2}q^{2}}{m^{2}}\Bigl(\frac{1}{A} - \frac{1}{A_{0}}\Bigr) + \frac{(1-h)q^{2}}{4}\SM (g_{0},g)
\\ +
\frac{2\pi \alpha_{q,\k}^{2}}{m^{2}}\Bigl(\frac{1}{A} - \frac{1}{A_{0}}\Bigr) +
\sum_{i}\bigl(2+k_{i}(k_{i}+q)\bigr)\sigma(x_{i}) \\
+ 2\pi\alpha_{q,\k}\sum_{i}k_{i}\bigl(\phi(x_{i}) - \SAY(g_{0},g)\bigr)
+ \frac{q\alpha_{q,\k}}{4}\SM(g_{0},g)\, .
\end{multline}
The first line in the above equation cancels out with similar contributions from the gravitational effective action when computing the gravitational expectation value in \eqref{gravexpdef}. According to \eqref{dressingCFT},
the term proportional to $\sigma$ in the second line provides the anomalous dimensions $\Delta(k) = -k(k+q)$ of the operators \eqref{opalpha} in the
massless theory. The new non-trivial part in the dressing factor is given by the terms in the third line of \eqref{dresslead}. Note that these terms do not contribute at $m=0$ simply because in this case the condition \eqref{CFTcond} must be imposed to obtain a non-zero correlator.

Subleading terms in the gravitational dressing, which do not
take a particularly simple or interesting form, can also be derived from
the third line in equation \eqref{kdepsm}. The metric variations of the 
relevant terms can be easily deduced from the formulas
\begin{align}\label{lastvar} & G_{2}(x,y;g) = -\int_{\sigh}\!\d^{2}z
\sqrt{g(z)}\, G(x,z;g)G(y,z;g)\\ \label{lastvar2}&
\int_{\sigh}\!\d^{2}y\sqrt{g(y)}\, G_{2}(x,y;g) R(y)  = -
\int_{\sigh}\!\d^{2}y\sqrt{g(y)}\, G(x,y;g)\psi(y;g)
\end{align}
together with \eqref{PsiAY} and \eqref{psiricdiff}.

\subsection{The stress-energy tensor}
\label{stresssec}

The results of the previous subsection immediately yield the 
trace $t(x)$ of the stress-energy tensor for our model, by computing 
the infinitesimal variations of the effective gravitational actions,
\be\label{tgravdef} \delta_{g}\Sgrav(g_{0},g) = \frac{1}{2\pi}
\ish t\delta\sigma\, .\ee
For example, from \eqref{tracearea}, \eqref{traceLiou} and \eqref{tMform} we find that
\be\label{tgrava} t(x;q,m) = \frac{8\pi^{2}(1-h)^{2}q^{2}}{m^{2}A^{2}}
+ \frac{2\pi}{A}\bigl(1 + (1-h)q^{2}\bigr)
+ \frac{25-3 q^{2}}{12} R(x) +\frac{2\pi(1-h)q^{2}}{A}\psi (x)
\ee
for the gravitational action \eqref{sleadq}. The term proportional to the Ricci curvature is the standard conformal anomaly, whereas the new term
proportional to the Ricci potential comes from the Mabuchi contribution
in the gravitational action. Similarly, for the gravitational action 
\eqref{sleadm}, we find, using in particular \eqref{dAY},
\be\label{tgravb} t(x;m) = \frac{2\pi}{A} + \frac{25}{12} R(x)
+ m^{2} \bigl(\ln\ell - \ln 2 + \gamma\bigr) + \frac{1}{2}m^{2}\bigl(
1 + \psi(x) - 8\pi h u^{\text c}(x) + 4\pi\Psi_{\text G}(g)\bigr)\, ,
\ee
where $u^{\text c}(x;g)$ is uniquely defined by the equations
\begin{align}\label{ucdef1} & \Delta u^{\text c} = \frac{1}{A}-
\frac{\sqrt{g^{\text c}}}{\sqrt{g}}\, \cvp\\&
\label{ucdef2}\ish u^{\text c}= 0\, .\end{align}
Using \eqref{Riccipotdef1}, \eqref{Riccipotdef2} and \eqref{DGRgen}, we see that
\be\label{ucform}h u^{\text c}(x;g) = 
\frac{1}{2}\bigl( \Psi_{\text G}(g) - G_{R}(x;g)\bigr) + \frac{\psi(x;g)}{8\pi}\ee
and thus
\be\label{tgravb2} t(x;m) = \frac{2\pi}{A} + \frac{25}{12} R(x)
+ m^{2} \bigl(\ln\ell - \ln 2 + \gamma\bigr) + \frac{1}{2}m^{2}\bigl(
1 + 4\pi G_{R}(x)\bigr)\, .
\ee

Let us note that the constant and metric-independent terms in $t(x)$ are 
scheme-dependent and can be absorbed by redefining the cosmological constant. Here we have indicated the precise results in the $\zeta$-function renormalization scheme. All the other terms are of course scheme-independent.

The trace of the stress-energy tensor could also be used to \emph{derive} the effective gravitational action, by proceeding backward.
The classical stress-energy for the action \eqref{dilatongen} is
\be\label{Tdil} T_{\text{cl}}^{ab}= 
-\frac{1}{2}\partial^{a}X\partial^{b}X + 
\frac{1}{4}g^{ab}\partial_{c}X\partial^{c}X + \frac{q}{2}\bigl(
\nabla^{a}\partial^{b}X - g^{ab}\nabla^{c}\partial_{c}X\bigr)
+\frac{m^{2}}{4}X^{2}g^{ab}\, ,\ee
and its trace is simply
\be\label{traceT} t_{\text{cl}} = \frac{q}{2}\Delta X + \frac{1}{2}m^{2}X^{2}\, .\ee
The quantum trace is then given by
\be\label{tq1} t = \frac{q}{2}\bigl\langle\Delta X\bigr\rangle_{\text{mat},\, g} + \frac{1}{2}m^{2}
\bigl\langle X^{2}\bigr\rangle_{\text{mat},\, g} + t_{\text{a}}(g)
\ee
in terms of expectation values in the matter theory and an anomalous
term $t_{\text{a}}$ that can be interpreted as coming from the variation of the path integral measure $\mathscr D X$ under Weyl transformations. From dimensional analysis, this anomaly must take the form
\be\label{tanoform} t_{\text{a}}(x) = a R(x) + M^{2}\, ,\ee
where $a$ is a dimensionless constant and $M$ some constant
mass scale. The parameter $a$ can be computed in the UV, i.e.\ in the massless scalar plus ghost CFT and it takes the familiar value $25/12$.
The mass scale $M$ can be absorbed in a redefinition of the cosmological constant and is scheme-dependent. In the $\zeta$-function scheme, the only mass parameter in the model is $m$ and thus we must have
\be\label{tanoform2} t_{\text{a}}(x) = \frac{25}{12} R(x) + b m^{2}\, ,\ee
for some number $b$. We shall see below that $b=1/2$.

The expectation values in \eqref{tq1} can be straightforwardly evaluated by computing the relevant gaussian path integrals,
\begin{align}\label{DeltaXvev} 
\bigl\langle\Delta X(x)\bigr\rangle_{\text{mat},\, g} & = -\frac{q}{2}R(x)
+\frac{1}{2}qm^{2}\int_{\sigh}\!\d^{2}y\sqrt{g(y)}\, G(x,y;m;g) R(y)\, ,\\
\label{X2vev} \bigl\langle X^{2}\bigr\rangle_{\text{mat},\, g} &=
4\pi G_{R}^{(\zeta)}(x;m;g) + \frac{q^{2}}{4}\biggl[
\int_{\sigh}\!\d^{2}y\sqrt{g(y)}\, G(x,y;m;g) R(y)\biggr]^{2}\, .
\end{align}
In the $\zeta$-function renormalization scheme, we have replaced the self-contraction of $X$ by the
function $G_{R}^{(\zeta)}$ defined in \eqref{GRzetadef} and related to 
$G_{R}$ by the relation \eqref{GRGRrel}. The small mass expansions \eqref{smallmG}, \eqref{smallmGR} and equation \eqref{lastvar2} then yield
\begin{multline}\label{tdirectm2} t(x) = 
\frac{2\pi}{A}\bigl(1 + (1-h)q^{2}\bigr) + 
\frac{8\pi^{2}(1-h)^{2}q^{2}}{m^{2}A^{2}} + b m^{2} \\ + 
\frac{25-3 q^{2}}{12}R(x) + \frac{2\pi (1-h) q^{2}}{A}\psi(x) 
+ 2\pi m^{2} G_{R}^{(\zeta)}(x) \\ + 
\frac{1}{8} q^{2}m^{2}\psi(x)\bigl(\psi(x) + 2\bigr) -
\frac{2\pi (1-h) q^{2}m^{2}}{A}\int_{\sigh}\!\d^{2}y\sqrt{g(y)}\,
G(x,y)\psi(y) + \mathcal O\bigl(m^{4}\bigr)\, .
\end{multline}
From this equation, we immediately find the leading non-trivial corrections to the CFT case. When $q\not = 0$, we can neglect the terms of order $m^{2}$ and we reproduce \eqref{tgrava}. When $q=0$, we find a match with \eqref{tgravb2}, taking into account \eqref{GRGRrel} and with $b=1/2$. In the most general case, it is straightforward to check using \eqref{psiricdiff} that the variation
\be\label{varlast} \frac{q^{2}m^{2}}{32\pi}\delta\ish \psi^{2}\ee
of the contribution \eqref{psilevcont} to the effective action precisely yields the terms in the third line of \eqref{tdirectm2} in the 
stress-energy.

Of course, the same method could also be used to find again the gravitational dressing factors, in particular formula \eqref{dresslead}, by including into the tree-level action the terms associated with the insertion of the operators, as in \eqref{Zdil}. This yields the additional contribution
\be\label{addopstress} 
-2\pi \sum_{i}k_{i}^{2}\frac{\delta(x-x_{i})}{\sqrt{g}}\ee
to the trace of the stress-energy and modifies in a rather
straightforward way the expectation values $\langle\Delta X(x)\rangle_{\text{mat},\, g}$ and $\langle X^{2}\rangle_{\text{mat},\, g}$. We let the details to the reader.

\section{Conclusion}
\label{concsec}

Studies of two-dimensional gravity have focused almost entirely on the Liouville model. This model is singled out since it universally describes 
the coupling of a CFT to gravity. Our aim in the present work was to motivate the study of different models, based on different actions like the Mabuchi functional, that are singled out by their nice geometrical features 
and their fundamental importance in K\"ahler geometry. Our main result
was to show that these models do appear naturally in simple examples, like the massive scalar field theory studied in our work.

We believe that the study of these new quantum gravity models could be of great theoretical interest. There are many natural questions one would like to answer. For example: how does the addition of the Mabuchi action
to a standard Liouville model modify the properties of the random surfaces?
How does it modify the usual KPZ relation \cite{KPZ}?
Does it help in going
through the $c=1$ barrier? What are the properties of a pure Mabuchi theory?... The new framework for the theory
of two-dimensional random surfaces recently developped in \cite{FKZ1} and \cite{FKZ2} seems well adapted to try to address these questions, either analytically or numerically.

Many generalizations along the lines of the calculations that we have
presented above can also be considered. Let us mention in particular the possibility to construct supersymmetric versions of the Mabuchi action and its generalizations by coupling to supergravity the supersymmetric version of the 
massive scalar field. Unravelling the structure of these new supersymmetric actions and studying the possible supersymmetric extensions of the higher dimensional versions of the Mabuchi action as well, are examples of intriguing natural questions one would like to investigate, with potential applications both in quantum gravity and in K\"ahler geometry.

\subsection*{Acknowledgements}

This work is supported in part by the belgian Fonds de la Recherche
Fondamentale Collective (grant 2.4655.07), the belgian Institut
Interuniversitaire des Sciences Nucl\'eaires (grant 4.4505.86), the
Interuniversity Attraction Poles Programme (Belgian Science Policy), the russian RFFI grant 11-01-00962 and the american NSF grant DMS-0904252.

\renewcommand{\thesection}{\Alph{section}}
\renewcommand{\thesubsection}{\arabic{subsection}}
\renewcommand{\theequation}{A.\arabic{equation}}
\setcounter{section}{0}

\section{On the derivatives of the Green's function}

Our aim in this Appendix is to provide an elementary derivation of 
\eqref{mixedpart}. We use a coordinate system in which, locally, the metric takes the form \eqref{metscoo}. It is convenient to use the complex coordinates
\be\label{uvdef} u=x^{1}+ix^{2}\, ,\quad v=y^{1}+iy^{2}\ee
and the associated partial derivatives. We are going to show that
\be\label{pderG} \partial_{u}\partial_{\bar v}G(x,y;g) =
\frac{1}{4}\delta (x-y) - \frac{1}{4}(\im\tau)^{-1}_{ij}l_{i}(u)\bar l_{j}(\bar v)\, .\ee
Since $\partial_{1a}\partial_{2a} = 2(\partial_{u}\partial_{\bar v}
+\partial_{\bar u}\partial_{v})$, the equation \eqref{mixedpart} used in the main text follows immediately from \eqref{pderG} in the limit $x\rightarrow y$.

The simplest way to understand \eqref{pderG} is to interpret it as a completeness relation in the space $\Omega^{(1,0)}(\sigh)$ of $(1,0)$ differential forms on $\sigh$, endowed with the positive-definite hermitian scalar product
\be\label{hilb} \langle\eta|\chi\rangle = -\frac{i}{2}\int_{\sigh}\bar\eta
\wedge\chi\, .\ee
The completeness relation follows from the the following simple special case of Hodge's decomposition theorem:

\noindent Lemma: \emph{The space $\Omega^{(1,0)}(\sigh)$ decomposes into a direct sum}
\be\label{directsum} \Omega^{(1,0)}(\sigh) = D\oplus H\, ,\ee
\emph{where $H$ is the space of holomorphic one-forms on $\sigh$ and $D$ is the set of $(1,0)$ forms which can be written as $\partial f$ for some function $f$ on $\sigh$. Moreover, the spaces $D$ and $H$ are orthogonal with respect to the scalar product \eqref{hilb}.}

The proof goes as follows. For any $(1,0)$ form $\eta$, $\partial\eta = 0$, which implies
\be\label{intonezero} \int_{\sigh}\bar\partial\eta = \int_{\sigh}\d\eta = 0\, .\ee
By the $\partial\bar\partial$-lemma we can thus write $\bar\partial\eta = \bar\partial\partial f$ for some function $f$ on $\sigh$. Indeed, in terms of components, $f$ satisfies
\be\label{deltafbarbar} \Delta f = -\frac{4}{\sqrt{g}}\partial_{\bar u}\eta_{u}\, ,\ee
and since the integral of the right-hand side is zero this equation always have a solution for $f$, given explicitly in terms of the Green's function as
\be\label{solgreenexpl} f(x) = 2i\int_{\sigh}G(x,y;g)\bar\partial\eta (y)= -2i\int_{\sigh}\bar\partial_{2}G(x,y;g)\wedge\eta\, ,\ee
where the Dolbeault operator $\bar\partial_{2}$ acts on the second argument of $G$. By construction, $\bar\partial(\eta - \partial f) = 0$ and thus
\be\label{etadec3} \eta = \partial f + \chi\ee
where $\chi$ is a homolorphic one-form. This is equivalent to the decomposition \eqref{directsum},
the orthogonality of $D$ and $H$ simply coming from the fact that, if
$\chi \in H$ and $\xi=\partial f\in D$,
\be\label{derperp1} \langle\chi|\xi\rangle = \frac{i}{2}\int_{\sigh}\d(
f\bar\chi) = 0\, .\ee

The matrix of the scalar product \eqref{hilb} on $H$ in the basis $(\lambda_{i})$ satisfying \eqref{onefnorm} and \eqref{period} is $\im\tau$ and thus the orthogonal projector $\Pi$ on $H$ is given explicitly by
\be\label{piformula}\Pi (\eta) = (\im\tau)^{-1}_{ij}\langle\lambda_{j}|\eta\rangle \lambda_{i}\, .\ee
In the decomposition \eqref{etadec3}, one has $\chi = \Pi(\eta)$. Taking into account \eqref{solgreenexpl} and \eqref{piformula}, Hodge's decomposition of the $(1,0)$-form $\eta$ is thus explicitly given by
\be\label{Hodgeexplicit} \eta (x)=-2i\int_{\sigh}\Bigl(
\partial_{1}\bar\partial_{2}G(x,y;g) + \frac{1}{4}(\im\tau)^{-1}_{ij}
\lambda_{i}(x) \bar\lambda_{j}(y)\Bigr)\wedge\eta(y)\, , \ee
where the Dolbeault operator $\partial_{1}$ acts on the first argument 
of $G$. This equation, being valid for any $\eta$, 
is equivalent to the equation \eqref{pderG}, which completes the proof.

\renewcommand{\theequation}{B.\arabic{equation}}
\section{Green's function and $\zeta$-function regularization}

The aim of this Appendix is to prove \eqref{zetadschemes} or equivalently, using the fact that $\zeta(s)$ has a simple pole at $s=1$ with residue $A/(4\pi)$,
\be\label{zetap1}
\lim_{s\rightarrow 0}\Bigl[ \zeta(s+1) -\frac{A}{4\pi s}\Bigr] = A
\Bigl(\Psi_{\text G}(g) +\frac{1}{2\pi}\bigl(\ln\ell-\ln 2 + \gamma\bigr)
\Bigr)\, . \ee
More generally, we can give a definition of the renormalized Green's function at coincident points using the $\zeta$ function method as follows. We introduce the generalized kernel
\be\label{zetaged} \zeta(x,y;s) = \sum_{i>0}\frac{\psi_{i}(x)\psi_{i}(y)}{\lambda_{i}^{s}}\,\cvp\ee
defined in terms of the eigenfunctions $\psi_{i}$s and eigenvalues $\lambda_{i}>0$ of the laplacian. The ordinary $\zeta$ function \eqref{zetadef} is simply
\be\label{zetaGz} \zeta(s) = \ish\zeta(x,x;s)\ee
whereas the ordinary Green's function is simply
\be\label{Greenzetadef} G(x,y;g) = \zeta(x,y;1)\, .\ee
Subtracting the pole of $\zeta(x,x;s)$ at $s=1$ we can define
\be\label{GRzetadef} G_{R}^{(\zeta)}(x) =\lim_{s\rightarrow 1}\Bigl[
\zeta(x,x;s) - \frac{1}{4\pi}\frac{1}{s-1}\Bigr]\, .
\ee
We are going to prove that the two definitions \eqref{GRdef1} and \eqref{GRzetadef} are equivalent up to a constant shift:
\be\label{GRGRrel} G_{R}^{(\zeta)}(x) = G_{R}(x) +
\frac{1}{2\pi}\bigl(\ln\ell-\ln 2 + \gamma\bigr)\, .\ee
Equation \eqref{zetap1} follows from \eqref{GRGRrel} by integration.

The idea of the proof is to identify the part of the function $\zeta (x,y;s)$ which yields the singularity when $s\rightarrow 1$. This can be done using the integral representation
\be\label{intzeta} \zeta(x,y;s) = \frac{1}{\Gamma(s)}\int_{0}^{\infty}\!
\d t\, t^{s-1}\Bigl( K(x,y;t) - \frac{1}{A}\Bigr)\ee
in terms of the heat kernel
\be\label{hkdef} K(x,y;t) = \sum_{i\geq 0}e^{-\lambda_{i}t}\psi_{i}(x)
\psi_{i}(y)\ee
and the standard small $t$ expansion of the heat kernel,
\be\label{smalltHK} K(x,y;t) = e^{-d_{g}^{2}(x,y)/(4 t)}
\Bigl( \frac{1}{4\pi t} + \frac{R}{24\pi} + \mathcal O(t)\Bigr)\, .
\ee
The singularity at $s=1$ is due to the integration region around $t=0$ in \eqref{intzeta} and to the pole in $1/t$ in the expansion \eqref{smalltHK}. Thus, the function defined by
\be\label{zetaRdef}\zeta_{R}(x,y;s) = \zeta(x,y;s)
- \frac{1}{\Gamma(s)}\int_{0}^{L^{2}}\frac{\d t}{4\pi}
e^{-d_{g}^{2}(x,y)/(4 t)} t^{s-2}\, ,\ee
where $L$ is an arbitrary length scale, is smooth at $s=1$. In particular,
\be\label{zetaRs1} \zeta_{R}(x,x;1) = \lim_{s\rightarrow 1}\zeta_{R}(x,x;s) = \lim_{y\rightarrow x}\zeta_{R}(x,y;1)\, .\ee
By evaluating
\be\label{zetxx} \zeta_{R}(x,x;s) = \zeta(x,x;s) - \frac{L^{2(s-1)}}{4\pi (s-1)\Gamma(s)}\ee
from the definition \eqref{zetaRdef} and using $\Gamma'(1) = -\gamma$, we find
\be\label{limze1} \lim_{s\rightarrow 1}\zeta_{R}(x,x;s) = 
G_{R}^{(\zeta)}(x) -\frac{1}{4\pi}\bigl(\gamma + \ln L^{2})\, .\ee
By similarly evaluating
\be\label{zets1ev} \zeta_{R}(x,y;1) = G(x,y) + \frac{1}{4\pi}
\text{Ei}\Bigl(-\frac{d_{g}(x,y)^{2}}{4 L^{2}}\Bigr)\ee
and using the expansion of the exponential integral function
\be\label{Eiexp} \text{Ei}(z) = -\int_{-z}^{\infty}\frac{e^{-t}}{t}\d t 
\underset{z\rightarrow 0^{-}}{=}= 
\gamma + \ln(-z) + \mathcal O(z^{2})
\ee
we find 
\be\label{zetRxyto} \lim_{y\rightarrow x}\zeta_{R}(x,y;1) = 
G_{R}(x) + \frac{1}{4\pi}\bigl(\gamma- 2 \ln 2 +\ln\frac{\ell^{2}}{L^{2}}
\bigr)\, .\ee
Equating \eqref{limze1} and \eqref{zetRxyto} as dictated by \eqref{zetaRs1} then yields \eqref{GRGRrel}, as was to be shown.

\end{document}